\documentclass[fleqn,usenatbib]{mnras}

\usepackage{newtxtext,newtxmath}

\usepackage[T1]{fontenc}

\DeclareRobustCommand{\VAN}[3]{#2}
\let\VANthebibliography\thebibliography
\def\thebibliography{\DeclareRobustCommand{\VAN}[3]{##3}\VANthebibliography}

\usepackage{graphicx}	
\usepackage{amsmath}	
\usepackage[export]{adjustbox}
\usepackage{xcolor}

\newcommand{\um}{\,\textmu{}m }	
\newcommand{\kms}{\,km s\textsuperscript{-1}}
\newcommand{\flux}{ erg s\textsuperscript{-1} cm\textsuperscript{-2} sr\textsuperscript{-1} cm}
\title[NGC 7538 IRS2 ]{NGC 7538 IRS2 in [NeII]:  Shell and Cavity Kinematics of a Compact HII Region}

\author[Dan Beilis]{Dan Beilis $^{1}$\thanks{danbeilis@mail.tau.ac.il}
Sara Beck,$^{1}$
John Lacy$^{2}$
\\
$^{1}$School of Physics and Astronomy, Tel Aviv University, Ramat Aviv, Israel 69978\\
$^{2}$Department of Astronomy, University of Texas, Austin Tx USA\\
}

\date{Accepted XXX. Received YYY; in original form ZZZ}

\pubyear{2022}

\begin{document}
\label{firstpage}
\pagerange{\pageref{firstpage}--\pageref{lastpage}}
\maketitle

\begin{abstract}
NGC 7538 IRS2 is a compact HII region and recent star formation source, with a shell morphology,  lying on the border of the visible HII region NGC 7538.   We present a spectral cube of the [NeII] $12.8$\um emission line obtained with the TEXES spectrometer on Gemini North with velocity resolution $\sim4$\kms~ and angular resolution $\sim0.3^{''}$.   The kinematics of the data cube show ionized gas flowing along multiple cavity walls.  We have simulated the kinematics and structure of IRS2 with a model of superimposed cavities created by outflows from embedded stars in a cloud with density gradients.  Most of the cavities, including the largest that dominates IRS2 structure, are associated with B-type stars; the outflow of the bright ionizing O star binary IRS2a/b is small in extent and lies in a high-density clump.  The IRS2 model shows that the behavior of an HII region is not a matter of only the most massive star present; cloud clumpiness and activity of lower mass stars may determine the structure and kinematics.  
\end{abstract}




\section{Introduction}

NGC 7538, 2.65 kpc distant, is an extended region of star formation. It holds a visible HII region,  on the western and southern edges of which are embedded star-forming complexes that excite strong infrared sources.  
The brightest are the massive embedded star-forming regions IRS1 and IRS2, shown in Fig.~\ref{fig:neii+radio_figure}, which are south of the visible source and obscured by $A$\textsubscript{V} at least 15 magnitudes.  IRS1 is an ultra-compact 
HII region excited by an O7 star \citep{sandell_2020}  at a very early stage of evolution. It is embedded in a dense disk (ibid.), which has collimated an ionized wind and is the site of methanol maser spots \citep*{minier_2001}, and drives a giant bipolar outflow of molecular gas  $\sim3$pc in extent. IRS2, about 10 arcsec north of IRS1, is a compact HII region about 0.13 pc in diameter. It has an overall
shell or cometary morphology and is a bright near and mid-infrared nebula.  \citet{zhu_2008} observed the HII region in the [NeII] 12.8\um line and report that with $\sim~2$ arcsec spatial resolution it has a cometary appearance and the kinematics of a pressure-driven outflow.  The exciting source, IRS2a/b,  is a close binary of a O9 and O5 star \citep{kraus_2006}. Near-infrared line images (\citet{bloomer_98},\citet{kraus_2006}) find shocked gas between IRS1 and IRS2 and multiple shells of gas centered on IRS2.  \citet{bloomer_98} suggest that these structures show the bow shock created as IRS2a/b moves relative to the cloud and to its own stellar wind.   

The close association and energetic activity of these embedded massive stars raise the question of the possible interactions with each other, with the rest of the young stars in the NGC 7538 region and with the gas clouds in their environment.   The molecular outflow of IRS1 appears to cover the IRS2 HII region; has it affected the development and appearance of the IRS2 nebula?  Has the stellar wind and bow shock of IRS2 any influence on the complex of outflows driven by IRS1?    To address these questions we must determine how the IRS2 HII region evolved into its present state, for which we need to understand the kinematics of the ionized gas.   While the molecular gas has been extensively observed with high spatial and spectral resolution, the ionized gas has not been as thoroughly studied: \citet{zhu_2008} had high (4\kms~) velocity resolution but relatively poor spatial resolution, with only 3-4 beams across the entire source, and \citet{bloomer_98} had better ($\sim1$ arcsec) spatial resolution but velocity resolution only $\sim350$\kms. The 21 cm HI observations of \citet{read_80} had formal velocity resolution $\sim4$\kms~ but the 2 arcmin spatial resolution was too low for useful information on the individual IRS sources.    

We have therefore observed the ionized gas in the IRS2 HII region with high spatial and spectral resolution, using the TEXES spectrometer on Gemini North to measure the 12.8\um [NeII] line with a sub-arcsec beam and true velocity resolution, including thermal broadening, of $\sim4$\kms~. Starting from the basic assumption that the HII region structure was created by stellar outflows in a dense cloud, we create, from PLUTO and the RADMC-3D simulations, models showing IRS2's evolution and iteratively compare them to the observations to arrive at the model most consistent with the data. In the next section we describe the observations and simulations and review data from the literature.

\section{Observations of The IRS2 HII Region} 
\subsection{ New [NeII] and Radio Maps}
The 12.8\um~[NeII] fine structure line is a useful  infrared tracer of gas ionized by young stars. [NeII] line images of compact and ultra-compact HII regions are seen to match very well with radio continuum images of equivalent angular resolution \citep{zhu_2008} except in extreme cases of very high local obscuration \citep*{Beilis_2021}.    In the density and ionization conditions of normal compact and ultra-compact HII regions essentially all the neon is singly ionized and emits this line.   While the exciting star of NGC 7538 IRS2 is unusually hot and may have possibly doubly-ionized some neon to $Ne^{++}$,    the spatial distribution of [NeII] in NGC 7538 agrees very well with the ionized hydrogen of the radio continuum maps (shown below). This argues that [NeIII] is not significant in IRS2 and  [NeII] can be relied on to trace the motions of the gas.   

\begin{figure}
		\includegraphics[width=0.95\columnwidth]{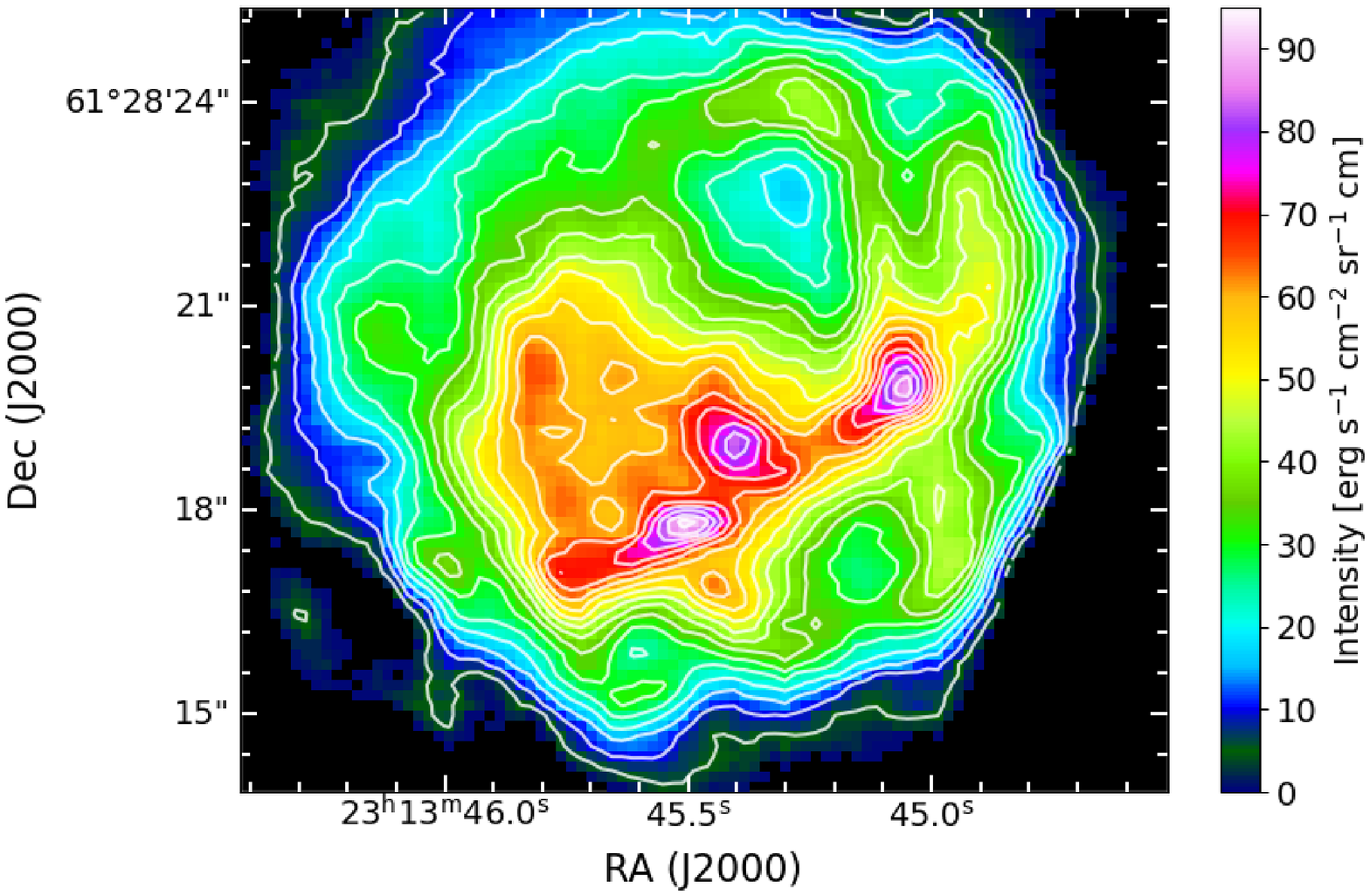}\\
		\includegraphics[scale=0.4]{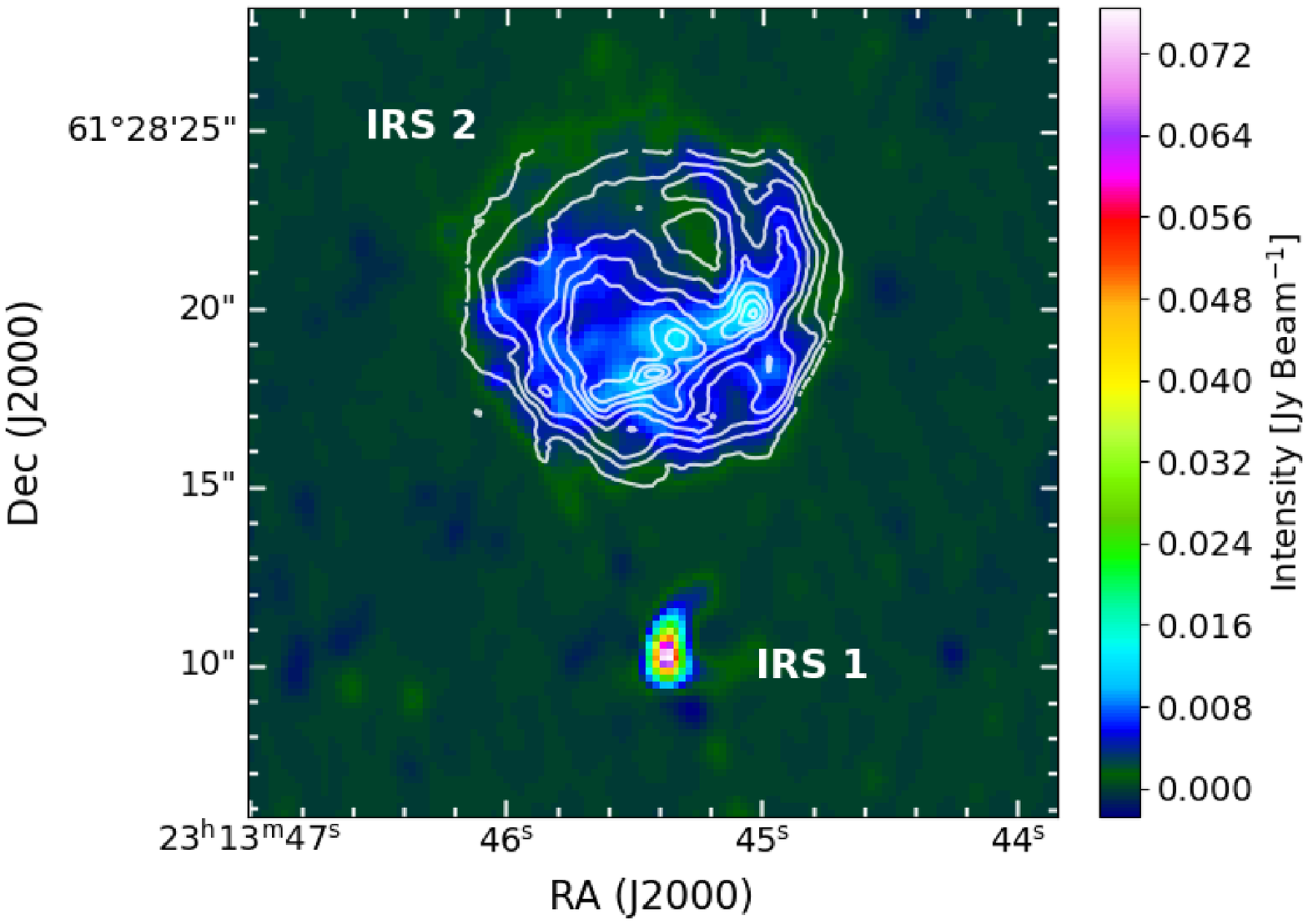}
    \caption{Top:  The zeroth moment of the sharpened [NeII] cube.  The noise level is $\sim1.8$ and the contour levels are 17 to 89.9 in steps of 9 (units\flux).   Bottom:
    the [NeII] contours superimposed on the 6 cm radio continuum map; the radio  noise level $\sim0.1$mJy bm$^{-1}$. }
    \label{fig:neii+radio_figure}
\end{figure}

We observed  the  [NeII] line in NGC 7538 at Gemini North on 7 November 2006  in program GN-2006A-DS-2 with the University of Texas echelle spectrograph TEXES \citep{lacy_2002}. TEXES is a high resolution spectrograph for wavelengths 5-25\um which uses a 256 $\times$ 256 element Raytheon Si:As array.  The diffraction and seeing limited beam size  at Gemini is  0.4 arcsec, corresponding  to $\sim 0.0045$ pc on the source, and the velocity resolution, including instrumental effects and thermal broadening, is $\sim 4 $\kms.  Each pixel along the N-S oriented slit was 0.15 arcsec; the slit was 0.59 arcsec wide and 4 arcsec long. The telescope was stepped $0.25 $ arcsec east-west across the source. 
The scans were merged to create a data cube with pixels 0.15 arcsec in declination $\times$ 0.314 arcsec in right ascension (square on the sky) $\times$ 0.92\kms~in velocity. The cube was further processed with a Maximum Entropy deconvolution; while this does not completely remove the point-spread function it sharpens the cube.   The registration of the cube was checked by comparison to an archival VLA map of the radio continuum at 6 cm with a $1.35\times1.09$~arcsec beam that was obtained in Program AP374 on 12 August 1998 and we conclude that the uncertainty in the absolute coordinates of the [NeII] cube is $\pm\sim0.2$~arcsec, a little more than 1 pixel.  Fig.~\ref{fig:neii+radio_figure} shows the total line emission (zeroth moment) map of the sharpened [NeII] cube and the [NeII] total superimposed on the 6 cm radio map.   (Both the radio and [NeII] maps are consistent with the [NeII] map of \citet{zhu_2008}, allowing for the greatly improved resolution of the current data). The spatial appearance of IRS2 is very similar in the two tracers, justifying the use of [NeII] as an ionized gas tracer.  The [NeII] observations did not cover IRS1, which appears in Fig.~\ref{fig:neii+radio_figure} as a bright radio source south of IRS2. 

\subsection{Data from the Literature }
\subsubsection{Stars}
\begin{figure}
\includegraphics[width=\columnwidth]{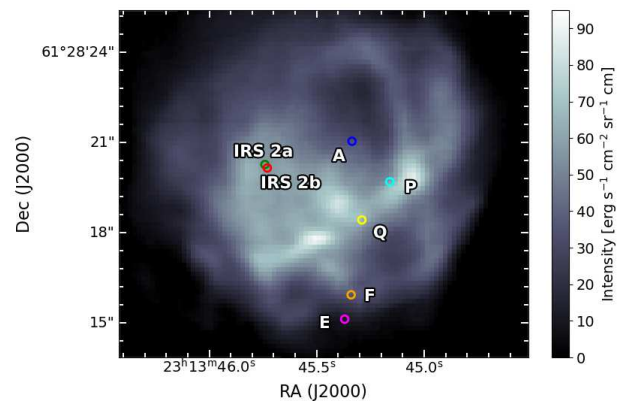}
    \caption{Locations of point sources measured by \citet{kraus_2006} on the moment 0 [NeII] map. The overlapping green and red circles show the two components of the main source IRS2a/b, the blue circle is the location of \citet{kraus_2006}'s star A, and the other circles correspond, from north to south, to their sources P,Q,F,E as labelled. }    
    \label{fig:stars_figure}
\end{figure}

\citet{kraus_2006} obtained  high resolution $K'$ band images of the IRS1 and IRS2 region with bispectrum speckle interferometry.  They found that the exciting star of IR2 is a close binary, IRS2a/b, and further located point sources. In Fig.~\ref{fig:stars_figure} we show the positions in the HII region of the stars and stellar candidates they report.   They find spectral types O5 and O9 for IRS2a and IRS2b respectively. From the relative brightnesses, the $K'$ magnitudes and \citet{ducati_2001} we estimate the spectral type of the brightest stellar candidates, stars A, P and Q, to be  B0-B1.5. The other point sources are almost 1 magnitude fainter than star A at $K'$ and their nature is unclear; they may be stars  of type B or later in NGC 7538 or non-stellar density peaks in the extended emission.  
\subsubsection{Molecular Clouds and Outflows}
The systemic velocity of the molecular cloud surrounding IRS2 and  IRS1 is  -57\kms~ \citep{Sandell_2009}.   The observed velocities of the molecular gas are dominated by the complex of powerful and 
very extended outflows driven most notably by IRS1,   The blue lobe of the molecular outflow extends over the position of IRS2 and has a velocity range between $\sim -80$ and $-64$~\kms~ (\citet{scoville_1986},\citet{sandell_2020}).    \citet{Sandell_2009} and \citet{bloomer_98} mapped shock tracers around IRS2 and suggested that the shock is excited from the south, by the IRS1 outflow. This interaction may also have created the complex layers or filaments of ionized gas apparent on the south edge of IRS2.  

\section{Kinematics of Ionized Gas in IRS2 -- Results} 
\subsection{Moment Maps}
\begin{figure}
\includegraphics[width=\columnwidth]{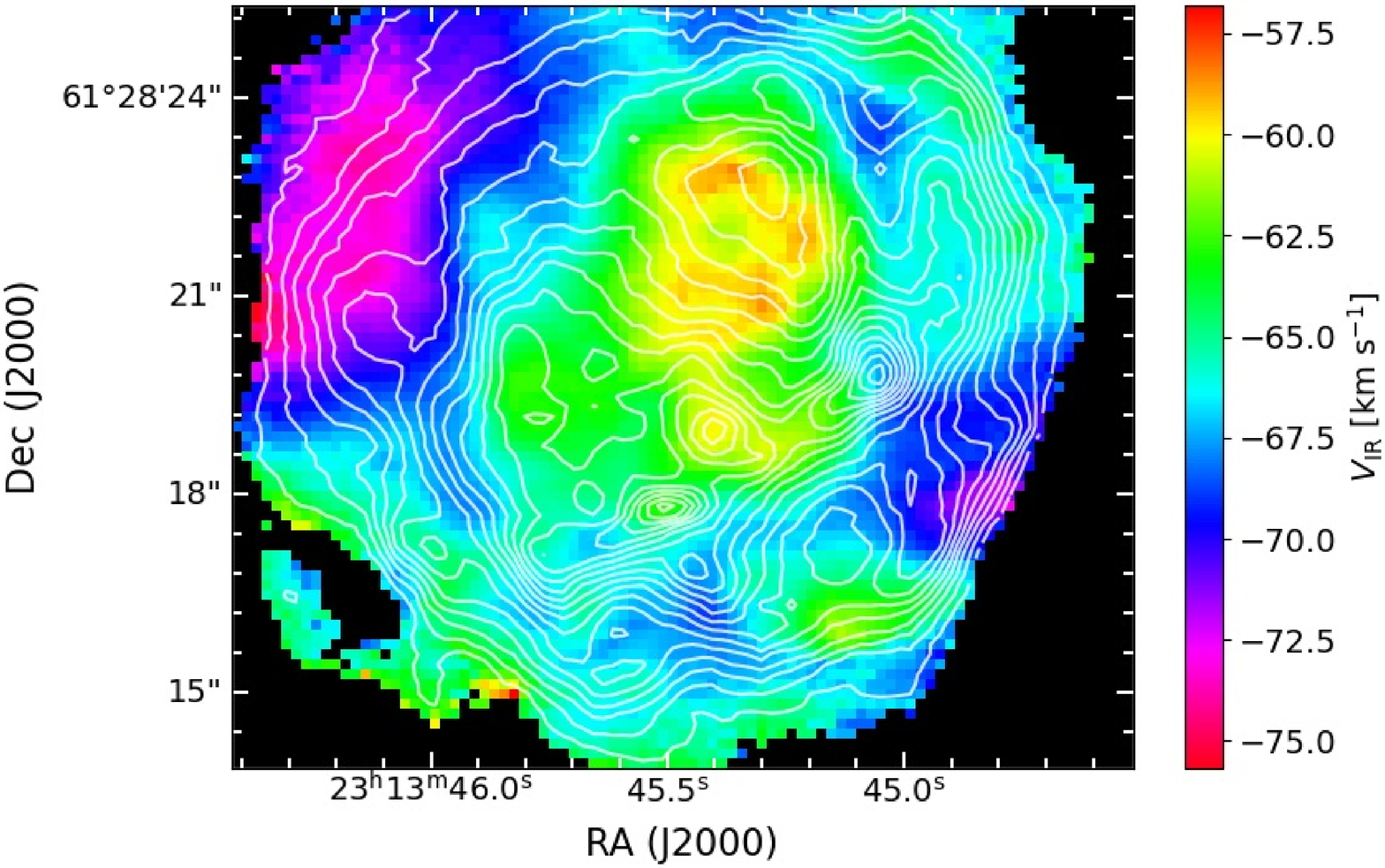}\\
\includegraphics[width=\columnwidth]{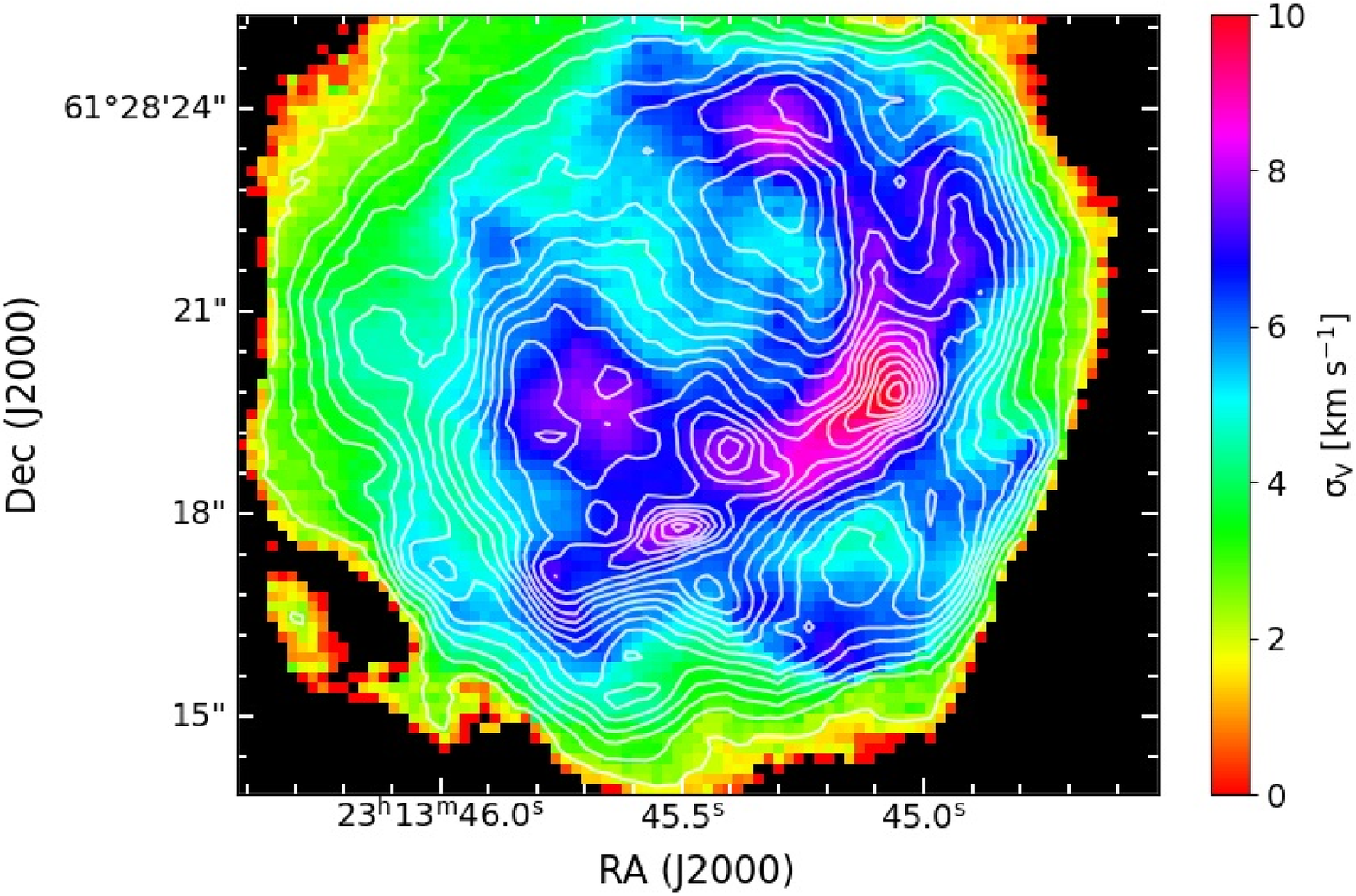}
    \caption{Top:  The first moment (mean velocity $V$\textsubscript{IR}) of the [NeII] cube. Bottom: The second moment $\sigma$\textsubscript{V} of the [NeII] cube. }    
    \label{fig:mom1+2_figure}
\end{figure}

Fig.~\ref{fig:mom1+2_figure} displays the first and second moments, respectively the intensity weighted velocity and the velocity dispersion, of the [NeII] data.  Both maps agree with the picture of overlapping shells or bubbles from the moment 0 and the radio maps.  The first moment map shows that the bulk of the [NeII] is significantly blue of the systemic cloud velocity of -57\kms, suggesting that the gas is expanding preferentially towards us, and that the rim is the most blue-shifted region.  The [NeII] line emission summed over the entire source, is close to a Gaussian centered at $-66$\kms and with $\sim20$\kms~ FWHM.  In the second moment map the dispersion is low and almost uniform on the rim of the HII region, while the central portion contains distinct areas of higher $\sigma$\textsubscript{V}, some of which will be shown in the next section to be affected by double-peaked line profiles.    The pattern of the dispersion shows that the gas has expanded more freely on the western side of IRS2 and suggests a shell  open on the west and with a closed vertex on the east.  The shell appears tilted into the plane of the sky. 

\subsection{ Position-Velocity Diagrams} 
 
\begin{figure}
\includegraphics[width=\columnwidth]{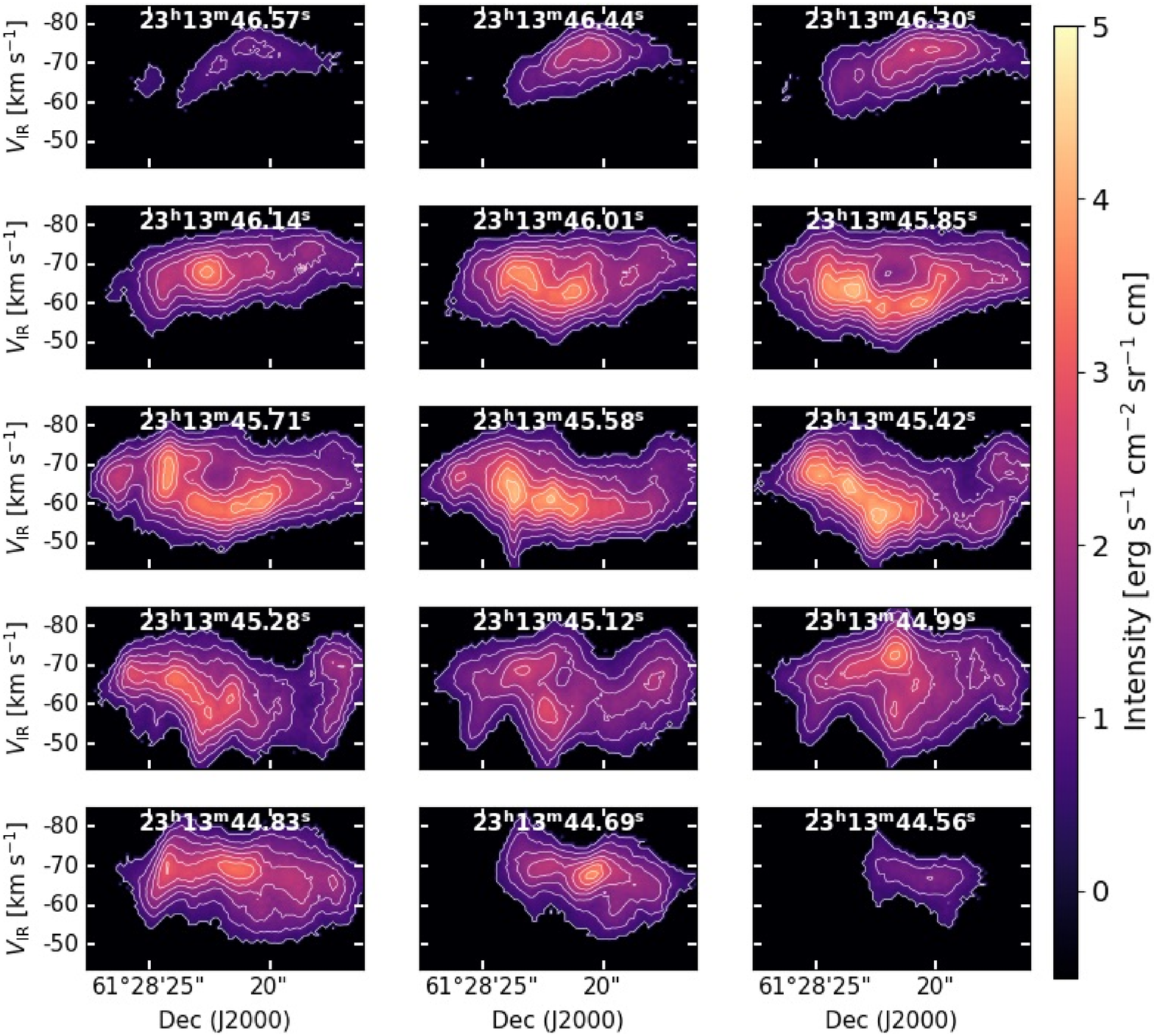} 

\vskip 0.2cm
\includegraphics[width=\columnwidth]{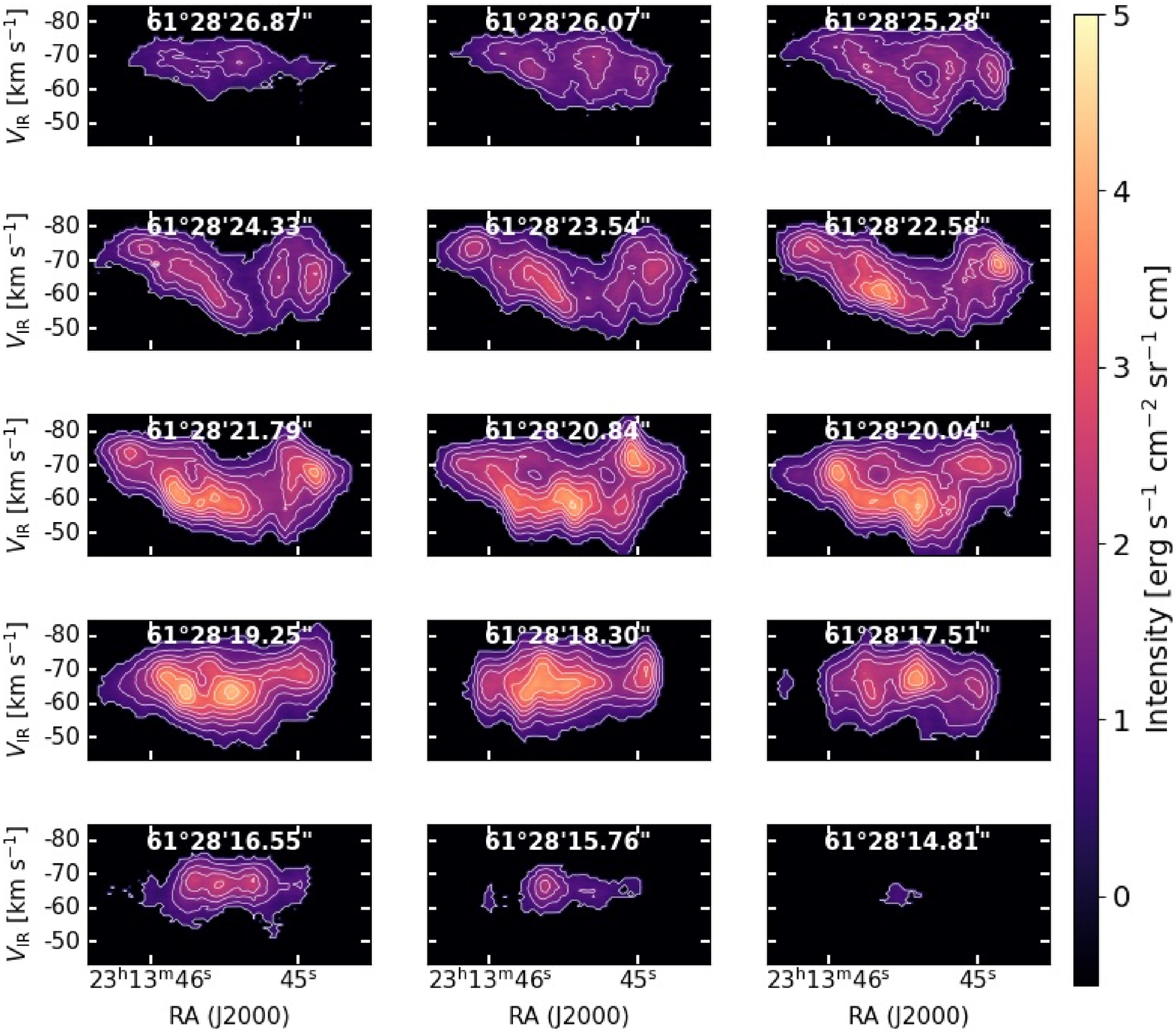}
    \caption{Position- Velocity Diagrams of the NGC 7538 data, showing the intensity of [NeII] emission at the given velocity $V$\textsubscript{IR} and position along cuts in Declination (top) and Right Ascension (bottom).  }    
    \label{fig:pvd_figure}
\end{figure}

 Fig.~\ref{fig:pvd_figure} shows the Position-Velocity Diagrams (PVDs) in cuts across IRS2 in Declination (top) and Right Ascension (bottom).  The width of the PVD (in the velocity dimension) shows the relatively small range of velocity across the source; the width of the line in one position is comparable to the maximum shift of the line peak across the source.   In many of the cuts the PVD is curved and clumpy,with the turn-around point in the curve close to the cloud velocity of $-57$\kms~ .  \citet{zhu_2008} and \citet{Immer_2014} display the PVDs produced by different types of gas flows. Comparing their models to our observations we see that the NGC 7538 PVDs closely resemble pressure-driven flows, with shell-like features in some positions.   We do not see the velocity offsets that characterize bow-shock systems in which the star moves into the cloud.  

\subsection{Working Picture: Multiple Overlapping Shells of Gas}
The observations of \citet{zhu_2008} with spatial resolution $\sim1.8$~arcsec led them to interpret IRS2 as a single limb-brightened source,  redder in the center than on the edge, and overall blue-shifted relative to the cloud. They assumed a single exciting star in the center and interpret the gas kinematics of IRS2 as a tangential outflow towards the observer along the walls of a stationary shell.   The much higher spatial resolution of our [NeII] data, together with current information on the stellar population,  gives a different and more complicated picture of the source.  

\begin{figure}
\includegraphics[width=1\columnwidth]{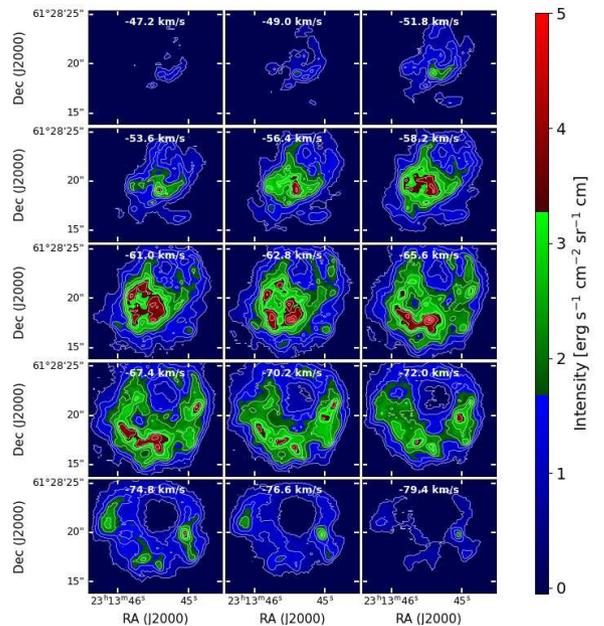} 
    \caption{Selected 0.92\kms~wide velocity channels of the [NeII] data cube; the velocities are given on each panel.  Contour levels are $n\times0.5$\flux$, n=1,2,3...10$. }
    \label{fig:channels_figure}
\end{figure}

\begin{center}
\begin{figure*}
\includegraphics[width=2.0\columnwidth]{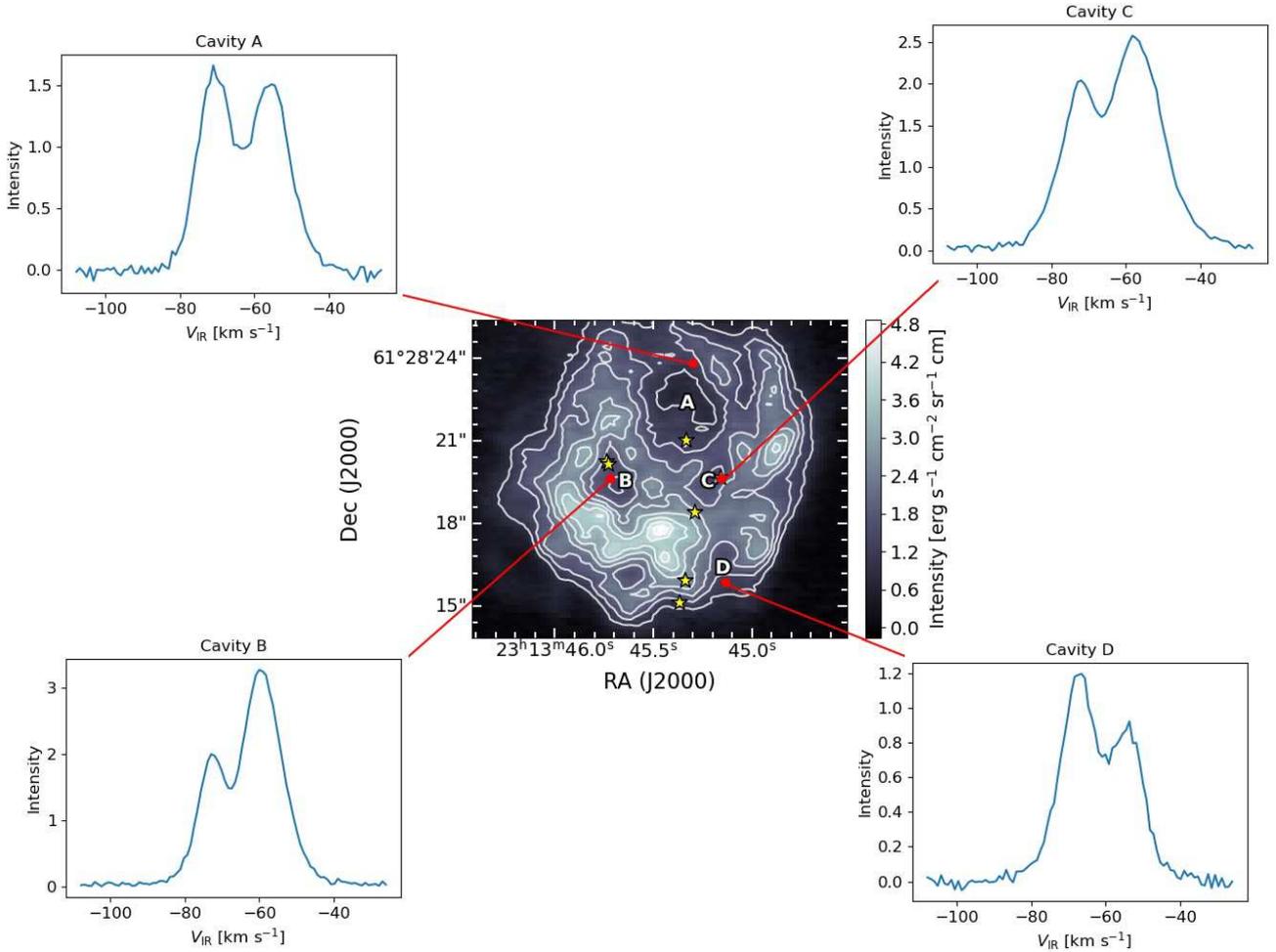}
\caption{ A map of the -65\kms~ velocity channel with the embedded stars and the 4 cavities marked and the [NeII] line profile at each cavity shown.  The red dots are approximately the beam size.   }
\label{fig:doublepeaks_figure}
\end{figure*}
\end{center}

First, the high spatial resolution in Fig.~\ref{fig:mom1+2_figure} shows several apparently overlapping shells and partial shells.   The  velocity dispersion is highest through the centers of the apparent shells (Fig.~\ref{fig:mom1+2_figure}).  This may suggest that the cavities are not empty but contain expanding lower-density gas, or, alternatively, that the cavities are not entirely open on the side facing us and that component of the gas flow adds to the apparent dispersion. 
Then, examining the velocity structure in detail with channel maps in Fig.~\ref{fig:channels_figure}, we see three main shells at slightly different velocities; the shells are identified and marked also in  Fig.~\ref{fig:doublepeaks_figure}.  The cavity marked 'A'  first appears at velocities close to the cloud velocity, is clear in the channel maps around $\sim-63$\kms~and persists through to the bluest velocities.  The velocity range of cavity 'B' overlaps with cavity 'A' on the red side;  cavity 'B' is clearest between $-62$ and $-67$\kms~ and fades out at more blue velocities, and the small cavities 'C' and 'D' first appear at velocities around $\sim-63$ and $\sim-73$\kms .  It should be noted that while the overall shift of the velocity blue relative to the cloud may suggest a cometary flow, there is not the relation of size and velocity that characterizes cometary flow features.  In a classic cometary flow (e.g. W33-Main-4, \citet*{Beilis_2021}) the  size of the feature increases with blue-shift. In NGC 7538 the source does appear to expand with velocity at the lowest velocities $\lessapprox65$\kms~, but the size is roughly constant between $-70$ and $-80$\kms~, and while it appears slightly larger in that range than at lower velocities, that may be an artifact of the higher S/N.  Alternatively,  the size-velocity correlation depends on the moving star having created a paraboloidal cavity and will not appear if the cavity is cylindrical instead. 

  The line profiles at every position on the source are shown in the Appendix Fig.~\ref{fig:profiles_figure}, and Fig.~\ref{fig:doublepeaks_figure} shows the spectrum in a 0.3~arcsec radius in each of the four candidate shells.    Fig.~\ref{fig:doublepeaks_figure}  and  Fig.~\ref{fig:mom1+2_figure} together demonstrate that the shells are the regions of highest velocity dispersion and that line profiles through the shells are double-peaked.  This is the spectral signature expected of expanding unfilled shells.   The spectra taken on the walls of the shells or cavities have single peaks with intensity higher than at positions in the shell; this agrees with the picture that the bulk of the gas flows along these walls. 

 \section{NGC 7538 Simulations and Model}
 \subsection{Motivation}
Our examination of the data cube in the previous sections shows that NGC 7538 IRS2 includes several cavities or shells and that the main cavities are associated with embedded stars or protostellar sources.   We now try to determine how this structure was created.   Where the low spatial resolution of previous studies showed a simplified picture of one large cavity and only one possible stellar source,  IRS2a/b,  we now have multiple cavities and potential driving sources to consider.     
 In this section we model the source as having been created by multiple stellar outflows expanding into a cloud with density gradients.  The positions of the cavities lead us to take as 
 driving stars IRS2a/b and \citet{kraus_2006}'s sources A,P,Q.   We have assigned stellar and wind parameters to each simulated star based on the spectral types as estimated in section 2.1.1 above, and the OB wind results of \citet{lamers1993mass}and \citet{krtivcka2014mass}.  The parameters are given in Table  \ref{tab:stellar_parameters}.  
 
It should be noted that young stars are active and known to drive many different forms of outflow and mass expulsion, ranging from highly collimated jets to high-velocity, low-density stellar winds.   Our models do not depend on, nor do we specify, any particular type or strength of activity. We assume only that some form of wind or outflow (here used interchangeably) was present and has created the observed cavities, along which the dense ionized gas (traced by [NeII]) flows, driven by pressure.

  We simulated the gas kinematics with PLUTO, as described in the next section, and iteratively adjusted the cloud density structure and stellar positions to best match the observed data cube.   In the next section we review PLUTO and the basic calculations, and in the following sections we discuss how we fine-tuned the stellar positions and density structure. 
\label{sec:maths} 

	\subsection{Hydrodynamics – PLUTO}
	We created the model of the HII region with the PLUTO hydrodynamics package \citep{mignone2007pluto}.  The simulation conditions and boundaries were as described in \citet*{Beilis_2021} and the cell width $4.8\times10^{-4}$ pc.  The boundary conditions in all directions were set to outflow. The Navier–Stokes equations of classical fluid dynamics were solved in an Eulerian method in 3D Cartesian coordinates using a HLL approximate Riemann solver \citep*{harten1983upstream}. $T_e=10^4$ K was assumed.  To simplify the model, we did not include heating and cooling processes; as the supersonic expansion of the gas into the evacuated bubbles should be close to isothermal \citep{franco1990formation} their effect should be small. Finally, a Courant number of $0.3$ was used based on the Courant-Friedrichs-Lewy condition \citep*{courant1967partial}.
	
	Each star starts with a spherically symmetric radial wind flowing from the origin of coordinates and remains stationary so that for an ambient medium density of $\rho\textsubscript{a}$ and pressure of $p\textsubscript{a}$, the initial conditions are: $\rho = \rho\textsubscript{a}$ and $p = p\textsubscript{a}$.
	The wind is injected within the internal boundary $r_0$ (inside the inner reverse shock of a stellar wind bubble) and based on \citet{mignone2014high} maintains these flow quantities constant in time:

	\begin{equation} 
		\setlength\arraycolsep{8pt}
		\begin{array}{cc} r^2 v_r \rho = r_0^2 V_0 \rho_0 & \text{(conservation of mass flux)} \end{array}
	\end{equation}
	\begin{equation} 
		\setlength\arraycolsep{8pt}
		\begin{array}{cc} v_r = V_0 \tanh\left(\frac{r}{r_0}\right)  &\text{(stellar wind acceleration structure)} \end{array}
	\end{equation}
	\begin{equation}
		\setlength\arraycolsep{8pt}
		\begin{array}{cc} p=\frac{c_s^2}{\Gamma} (\rho_0^{1-\Gamma}) \rho^\Gamma &\text{(pressure-density adiabatic relation)} \end{array}
	\end{equation}

	where $r$ is the radius, $\rho_0$ is the ambient density, $V_0$ is the gas velocity at $r_0$, $p$ is the gas pressure, $c$\textsubscript{s} is the sound speed of the gas and $\Gamma=c_P/c_V$ is the ratio of specific heat coefficients.   Values of $r_0$ and $V_0$ for the 4 simulated stars can be seen in Table \ref{tab:stellar_parameters}.

	\subsubsection{Locating Stars and IRS2a/b}
	The positions of the various stars in the X and Y axis (the Dec and RA axes in the plane of the sky) were based on the coordinates given by \citet{kraus_2006}, while the distances in the line-of-sight axis (Z-axis) were determined by a iterative fine-tuning process to give the best match to the observed data cube.   
	
The brightest stars in IRS2, and therefore the sources of the strongest winds, are in IRS2a/b: they are estimated to be type O9 and O5 \citep{kraus_2006} and their separation on the plane of the sky is only 0.195arcsec (ibid).  Will the interaction of the two star's outflows modify their effect on the ambient cloud?  As a test we simulated only those two stars in a constant ambient density.  We found that on length scales comparable to the distance between the stars, the interactions had significant effect, producing noticeable swirls and spirals.  But on length scales comparable to the distances between the embedded stars in IRS 2, the effects are negligible. The two stellar winds together produced a single minimally distorted spherical cavity, and the radius of the cavity was almost the same as what the more massive star (IRS2a) would make alone.   We therefore treat IRS2a/b in the simulations as a single star at the center of the binary's orbit and with IRS2a parameters.   

\subsubsection{Density Structure}
	The overall structure of IRS2 is dominated by the largest shell. It is blueshifted relative to the cloud and has PVD characteristic of a strong pressure driven flow \citep{zhu_2008}.  To match this in the model,  the density gradient includes both a power law function $H(r) \propto \rho^{-\alpha}$ and a step function.  This sharpens the $r\textsubscript{d}$ boundary between the low and high density areas \citep*{henney_2005}, and allows us to place the stars off-center 	at distance $\tilde{z}$ from the molecular cloud core.  We used a standard $\alpha=1.5$ \citep{pirogov_2009,sano_2010}.  $\rho_0$ was chosen to give density on the order of $10^5\text{ }$cm$^{-3}$ close to the edge of the high density region,  and $\tilde{\rho}_0 = 10^2\text{ }$cm$^{-3}$, so the ambient density profile is: 
	\begin{equation}		 
		\rho\textsubscript{a}(r,\tilde{z}) = 
		\begin{cases}
			\rho_0 H(r) & r < r\textsubscript{d} \\ 
			\tilde{\rho}_0 & r > r\textsubscript{d} \\
		\end{cases}
	\end{equation}
  The core's center was set at an approximately $45^{\circ}$ angle relative to the Z or line-of-sight axis, to match the PVDs.  IRS2a/b was placed at the edge of the high density area.   
  
  The density structure of the data cube indicate that stars A,P,Q are in regions of low constant ambient denstiy, 
  To fit this structure it is necessary to assume that these stars had already evacuated bubbles around themselves before the main IRS2a/b outflow started.   As they are B stars and much longer lived than the O stars that comprise IRS2a/b, this is realistic.  The density in the bubbles was set to $\rho\textsubscript{a} = 10\text{ }$cm$^{-3}$ and the sizes on the order of the observed shell sizes. We added a small low density bubble close to star P at the edge of star A's bubble, which reproduces a filamentary structure extending towards the observer in the data cube.  
	
	The region around IRS2a/b, Cavity B,  is particularly complex. The data cube indicates that a strong outflow has expanded unevenly into a clumpy and dense medium.  The channel maps of Fig.~\ref{fig:channels_figure} show that the south-west side of Cavity B is blue-shifted and its north-east edge relatively red-shifted. To match this we placed one small and two large low density ($\rho\textsubscript{a} = 10^2\text{ }$cm$^{-3}$) spheres near IRS2a/b. The large spheres north-east and south-west of IRS2a/b, are offset in the line-of-sight and are dominated respectively by red and blue outflowing material.
	
The proximity of the strong source IRS2a/b to the large cavity around star A suggests that the IRS2a/b outflow could expand into the low density Cavity A, but  simulations of the IRS2a/b and star A outflows interacting produced a wake which is not seen in the observations.  We therefore constrain the IRS2a/b outflow to remain within the high density Cavity B area.   We note that the mismatch with the data may be due to the limitations of the simulations and of the simple stellar outflow model that we have chosen, and does not rule out the possibility that IRS2a/b is also expanding into cavity A. 

\begin{figure}
		\centering
		\includegraphics[width=0.9\columnwidth]{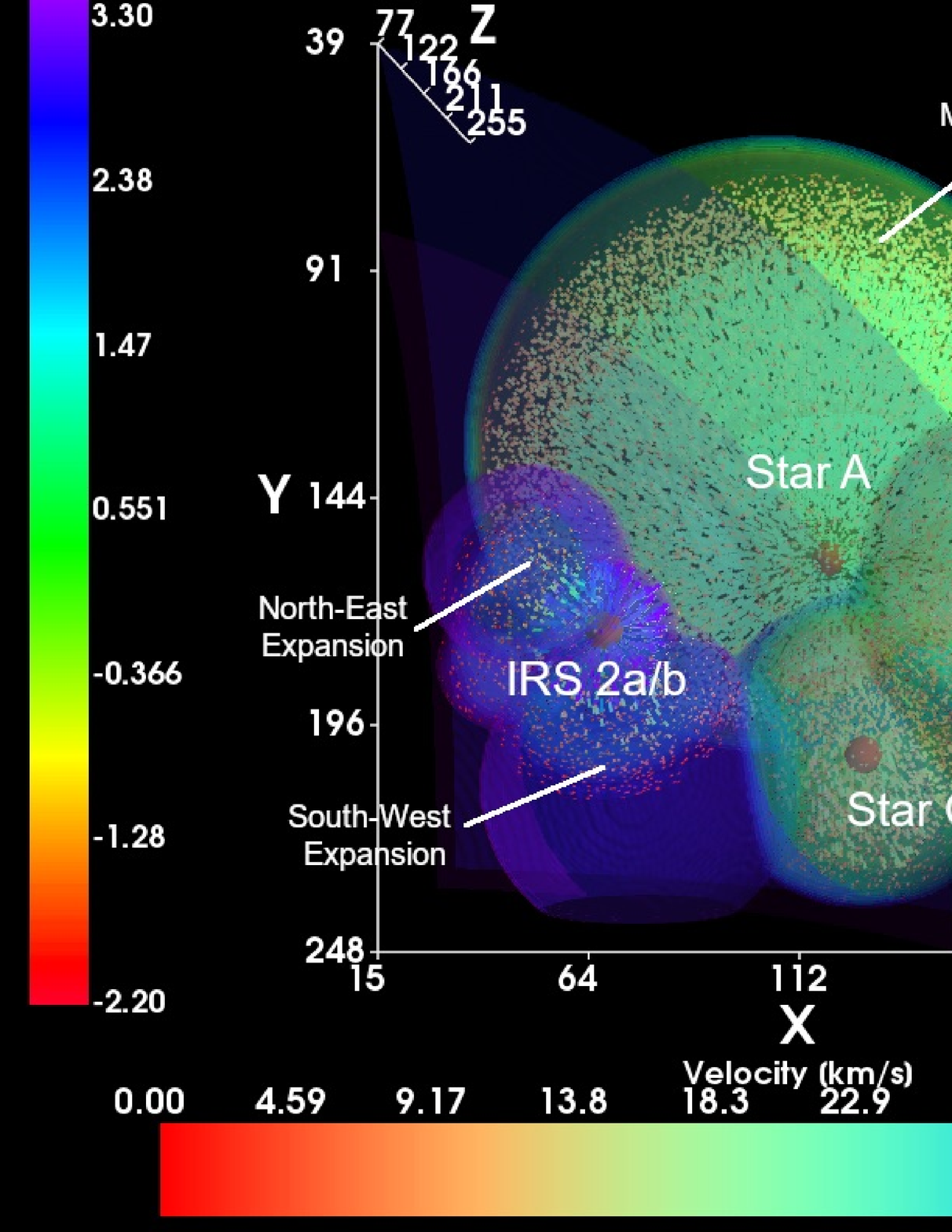}
		\includegraphics[width=0.9\columnwidth]{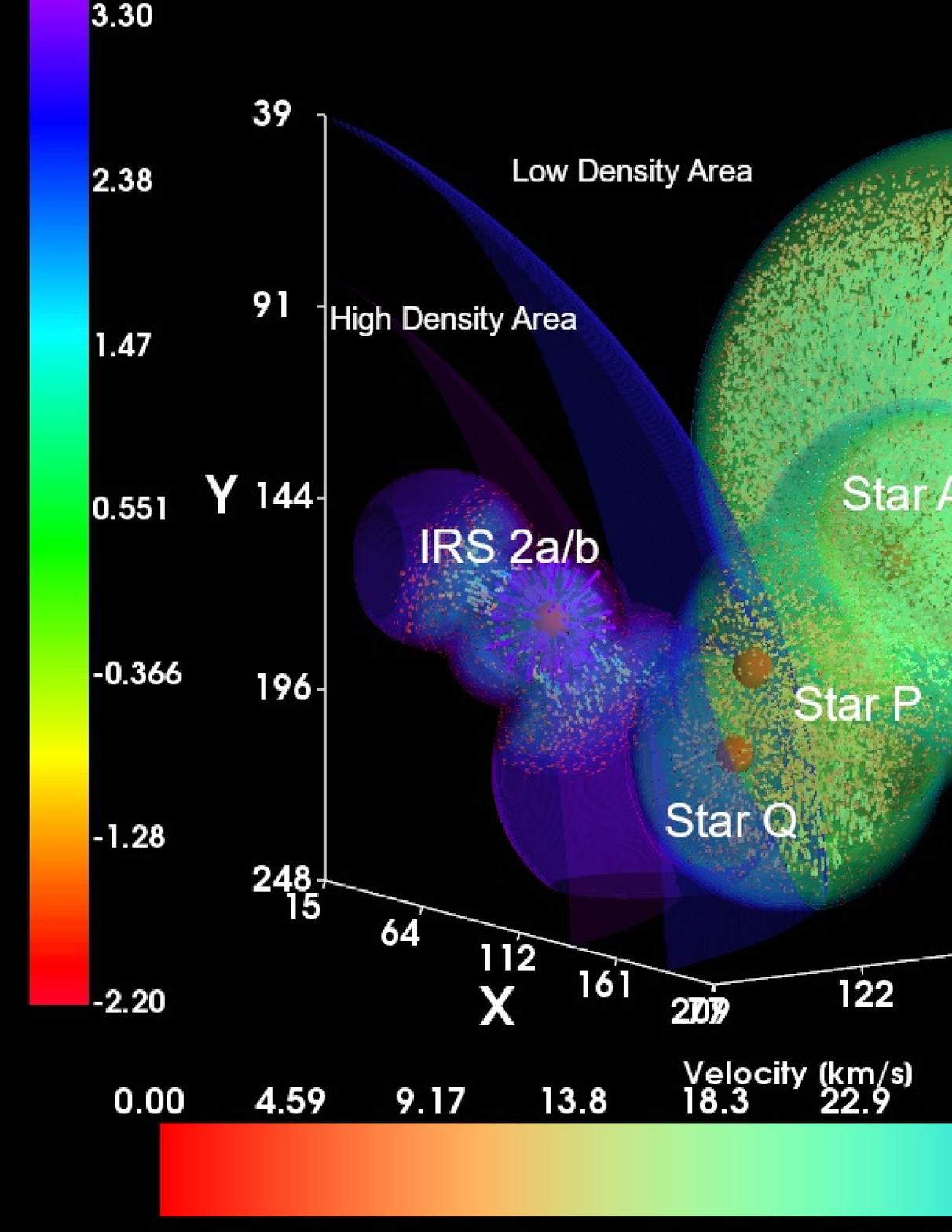}
		\includegraphics[width=0.9\columnwidth]{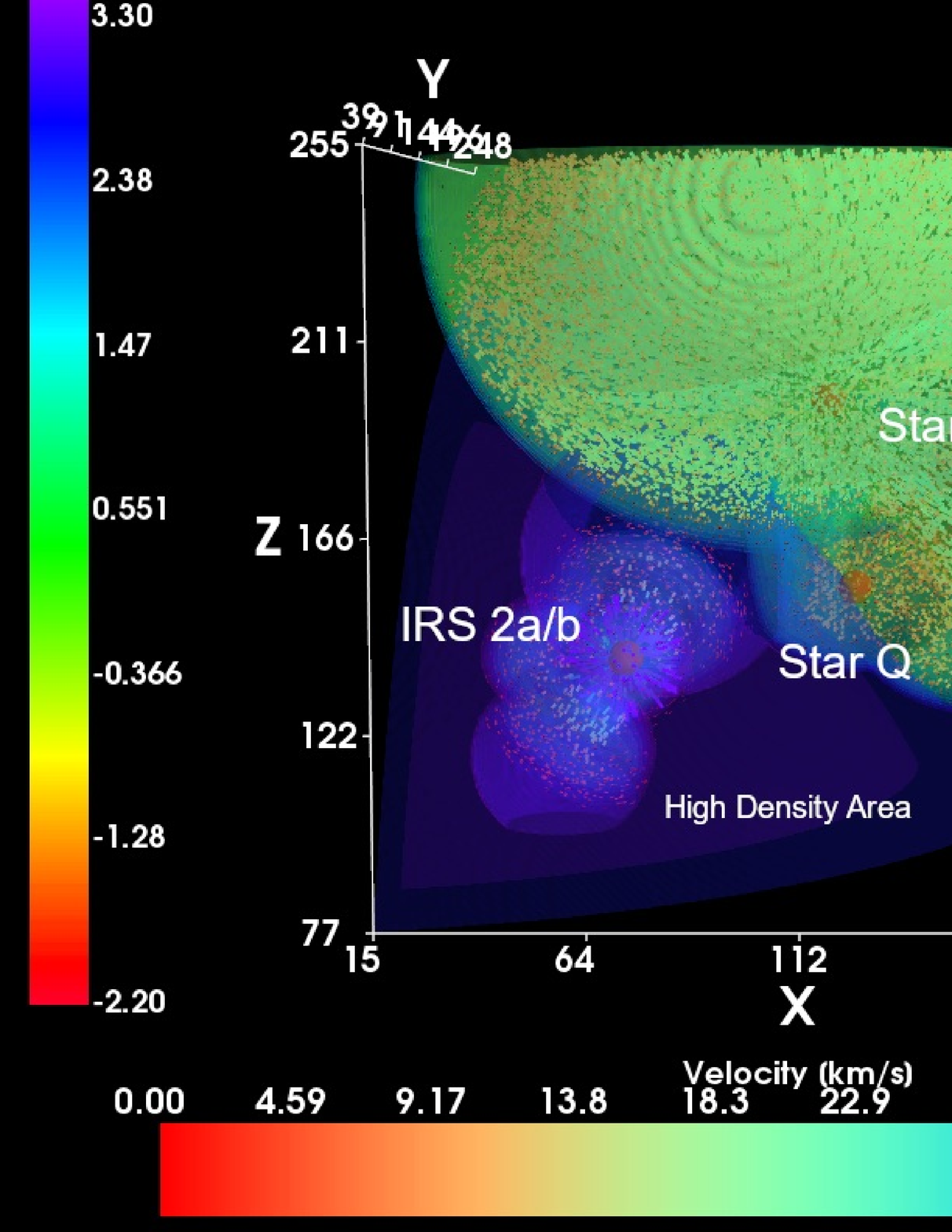}
		\caption{Multiple viewing angles of the [NeII] density model of NGC 7538 IRS 2 produced by PLUTO where the high and low density areas are marked on the figure and the stars are denoted by red spheres. 1 X,Y,Z unit is 0.00048 pc and the observer is located in the Z+ direction. The velocity scale was chosen for computational efficiency and is offset from the observed velocities by 80\kms~. \label{fig:hydro_model}}
	\end{figure}

 Fig.~\ref{fig:hydro_model} shows the most satisfactory model we have obtained.  In the figure the driving stars, the main cavities and the low and high-density structures are displayed from several angles; the full 3-dimensional model can be seen as a movie in the appendix.   Fig.~\ref{fig:model_moment0} shows the zeroth moment of the simulated cube, as viewed from our position.  
	
	\begin{table}
		\caption{Simulated stars' parameters}
		\label{tab:stellar_parameters}
		\begin{tabular}{lcccccc}
			\hline
			Star & Spectral Type & $M_* $ & $T_*\textsuperscript{eff}$ & $r_0$ & $V_0$\\
			& & M\textsubscript{\(\odot\)} & $10^4$ K & $10^{-3}$ pc & \kms\\
			\hline
			IRS2a/b & O5 & 51.00 & 4.43 & 6.49 & 40 \\
			A & B0V & 14.73 & 3.00 & 3.25 & 20 \\
			P & B1.5V & 9.26 & 2.40 & 1.91 & 12 \\
			Q & B1.5V & 9.26 & 2.40 & 1.91 & 12 \\
			\hline
			\multicolumn{5}{p{.8\columnwidth}}{Note: List of the simulated stars.   We have assigned to each star a mass ($M_*$), effective temperature ($T_*\textsuperscript{eff}$), internal boundary of the bubble created by the outflow ($r_0$), and $V_0$ the gas velocity at $r_0$, consistent with the spectral types as estimated in section 2.1.1.  }
		\end{tabular}
	\end{table}

	\subsection{3D ionic line emission profile simulation – RADMC-3D}
\begin{figure}
\includegraphics[width=1\columnwidth]{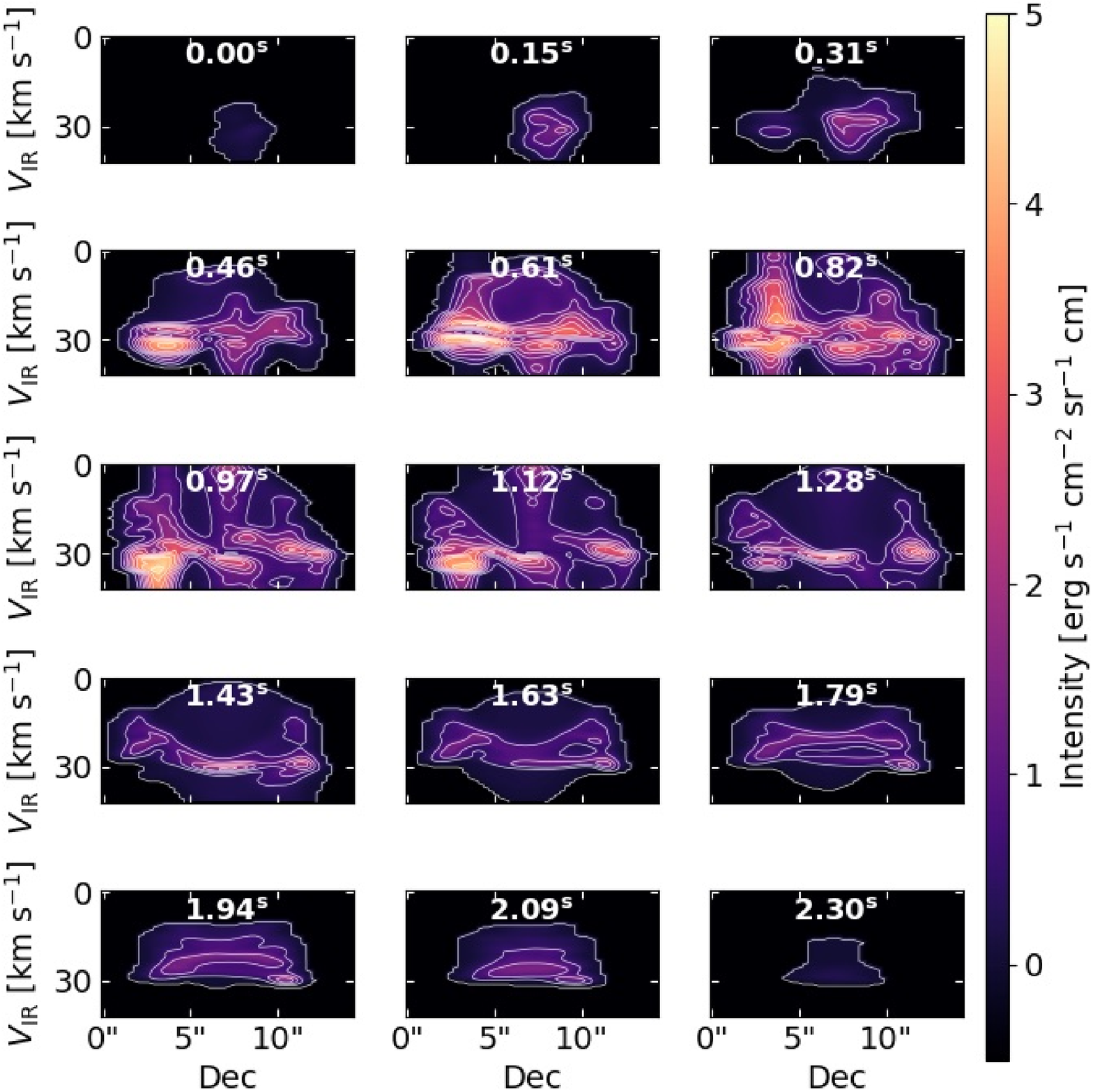}
\vskip 0.2cm
\includegraphics[width=1\columnwidth]{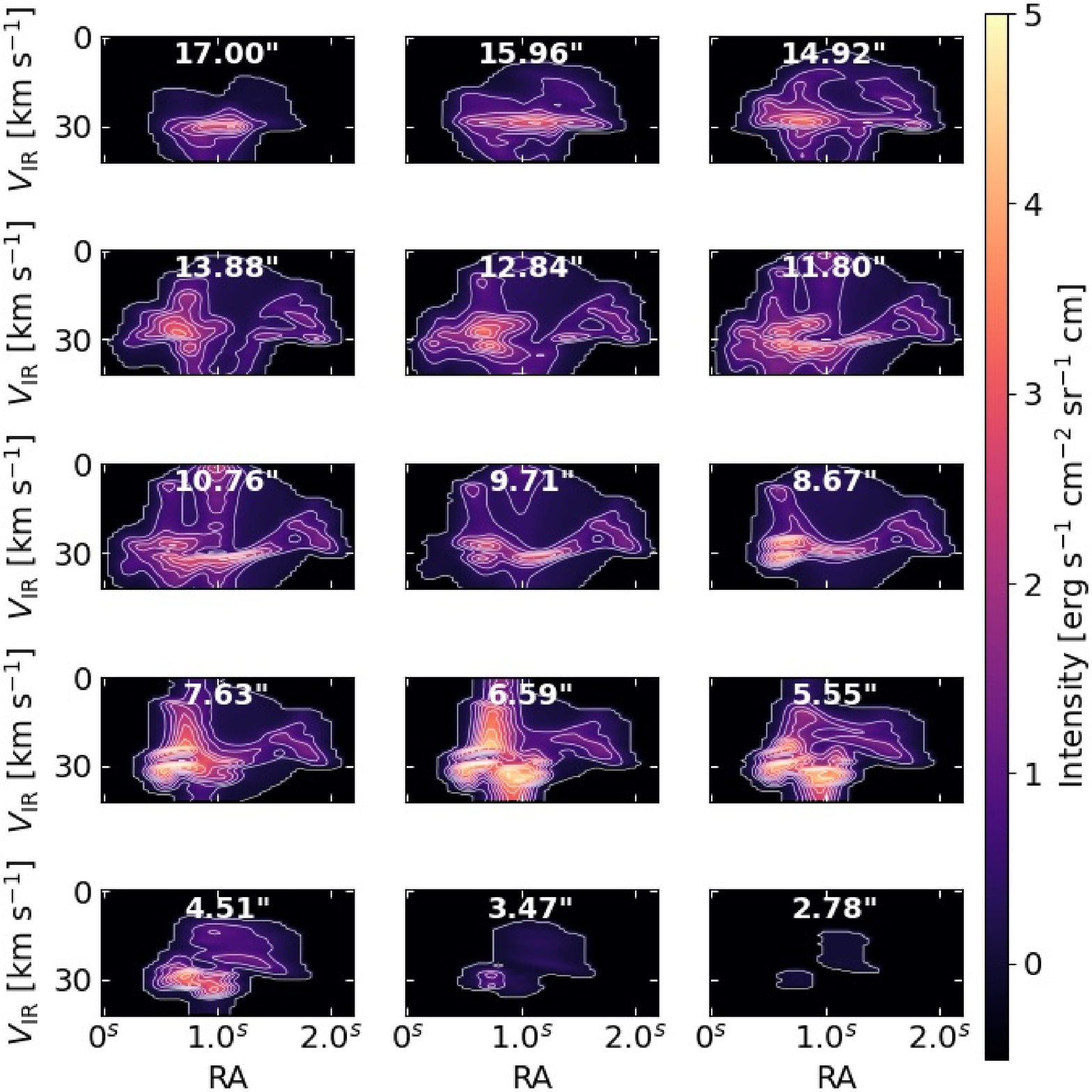}
\caption{Position-Velocity Diagrams showing the velocity of the simulated [NeII] 3D line profile at every position in the simulation results.  Cuts in Dec are at the top and R.A. on the bottom. The 0 of the velocity scale is arbitrary and the results convolved to match the Gemini $0.3''$ beam size.  }
\label{fig:pvdmodels_figure}
\end{figure}

	The PLUTO outputs (Ne+ density, velocity, and position) were input to the radiative transfer software RADMC-3D \citep{dullemond2012radmc}, which calculated the 3D [NeII] line profile emission. We assumed constant gas temperature $10^4$ K (the line intensity is not very sensitive to temperature, depending on $(\frac{10^4~K}{T_e})^{1/2}$).  The line profile was simulated at the spectral range and resolution matching the observed [NeII] data cube.    The PVDs extracted from the simulated cube are shown in Fig.~\ref{fig:pvdmodels_figure} and the channel maps in Fig.~\ref{fig:modelchannel_figure}.  	 (The velocity scale in the simulated cube was chosen for a computationally convenient zero, and therefore there is an overall shift in velocity between the simulated  observed data cubes).
	
	\begin{figure}
	\includegraphics[width=0.9\columnwidth]{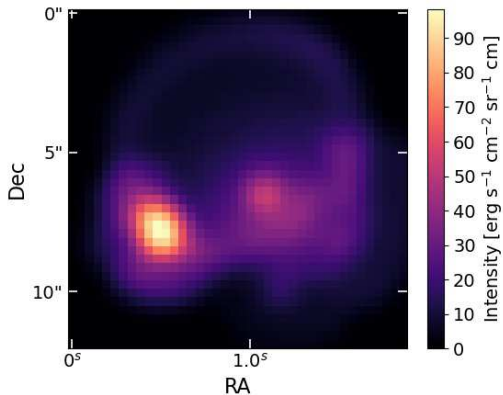}
	\caption{Moment 0 of the simulated data cube convolved with the Gemini beam size, showing total [NeII] emission.  \label{fig:model_moment0}}
	\end{figure}
	
		\subsubsection{Simulation Results and the Observed Data Cube}
		
		The moment 0 map of the simulated data cube agrees well with the observed spatial distribution of the radio continuum and the [NeII] (Fig.~\ref{fig:neii+radio_figure}).  A striking feature of the simulation is the ridge of bright emission running SE-NW.   In the simulation, the ridge is the overlap in the line of sight of the density features created by the outflows of IRS2a/b and stars A and P; the three intensity peaks show (east to west) the IRS2a/b outflow, the outflow of star A meeting the wall of Cavity A, and the Cavity A wall meeting the outflow of star~P.    
		
		This [NeII] ridge coincides spatially with the region in which \citet{bloomer_98} find emission from the shock tracers $H_2$ and [FeII]; they suggest that the shocks show where the stellar wind of  IRS2a/b impacts the molecular cloud. \citet{bloomer_98} further suggest that IRS2a/b has created a bow shock by moving through the cloud SE at $\sim10$\kms~.  The observed PVDs, however,  show the signature of pressure driven flows into cavities of stationary stars rather than the bow shock signature of moving stars.   The current data and the shock tracers may  be consistent if IRS2a/b is moving almost entirely in the plane of the sky, and so creating shocks but not the kinematic marker of a bow shock.  It is also possible that the shocks \citet{bloomer_98} observe were excited by the HII region expanding into the molecular cloud and do not reflect stellar motion.

\begin{figure}
\includegraphics[width=1\columnwidth]{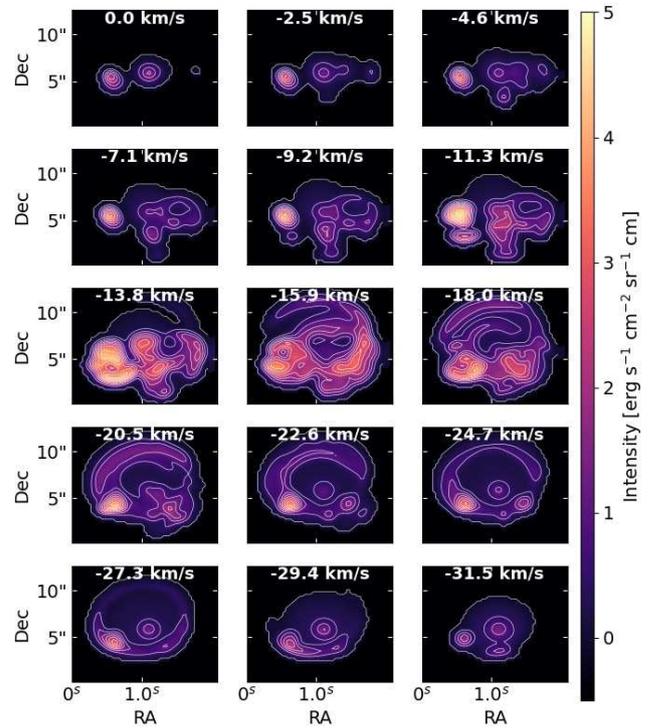}
\caption{Velocity channel maps from the simulated [NeII] 3D line profile,  convolved to match the  Gemini beam size. }
\label{fig:modelchannel_figure}
\end{figure}

	The channel maps of the observed (Fig.~\ref{fig:channels_figure}) and simulated (Fig. ~\ref{fig:modelchannel_figure} agree fairly well. In particular the simulations match the observed velocity development of Cavities A and B, giving us some confidence in our model for the complex setting of IRS2a/b.  The PVDs of the simulation match many but not all of the small brightness clumps in the original data, and confirm that the kinematics are champagne-type; that is,  with no 
	motion of the exciting star relative to the cloud.


\section{Discussion and Conclusions}
We have obtained high spatial and spectral resolution observations of the $12.8$\um [NeII] emission in NGC 7538 IRS2.  The ionized gas traced by the [NeII] line has formed several cavities or shells, and the PVDs suggest that the gas flows along the walls of cavities created by stellar outflows.   The observed cavities can be correlated with some of the young embedded stars observed in the near-infrared by \citet{kraus_2006}.   We have accordingly modelled the HII region as a collection of overlapping shells or bubbles created by outflows from the embedded stars.  Simulations of the gas kinematics from the model show: 
\begin{itemize}
\item  The observations can be well matched with 4 outflow cavities.  
\item   Three of the outflow cavities (A,B, and C) are associated with embedded stars; the fourth has no known stellar sources.  
\item  The early O binary pair IRS2a/b, the main luminosity and ionization source, is not associated with the largest cavity but is located in a small high-density region.
\item   The largest cavity, that dominates the source structure, is in a low-density region and holds an early B-type star.
 \item  The overlap of outflow cavities in the line of sight has created the bright clumpy ridge structure which dominates the continuum maps and which is thought 
 to be a shocked region.  
 \end{itemize}
 
 We do not have a timeline for the creation of the cavities, as we have no information about the strength or duration of the ouflows that formed them, nor about their relative ages.  We note that the O stars of IRS2a/b are short-lived compared to B-type stars, so stars A,P,Q may well be older than IRS2a/b and had longer to affect the cloud structure, and also that even complicated density structures may be formed quickly, especially if the ambient gas is not very dense.   In the conservative limiting case that one of the small cavities which appears as a closed sphere in the models is the result of a bubble expanding at $\sim5$\kms,  it would reach its current size of $\approx10^{-2}$ pc in only $2\times10^3$ years.

 The model we have developed gives results which agree with the observed data, but this does not necessarily mean that the model matches reality.   It is possible that the observed cavities were created by or influenced by some mechanism other than stellar outflows, or by stars other than the ones recorded by \citet{kraus_2006}.   The molecular medium around IRS2 is active and complex. The NGC 7538 region holds several embedded IR sources that drive large-scale molecular flows.  The closest to IRS2 on the plane of the sky is the very active source IRS1 which has a strong outflow.  It is possible, although not observed, that the IRS1 outflow has affected the IRS2 region. Observations of the molecular gas in the IRS2/IRS1 region with good spatial resolution could determine if the two sources are interacting; measurements of 
 shock tracers would likewise be useful.  
 
  NGC 7538 IRS2 is an excellent example of a mature compact HII region with multiple embedded stars.   Our data suggest that even though the ionization and luminosity of the source is dominated by one (binary) star,  the structure of the source has been shaped by the smaller stars; their outflows have created the cavities and shells in which the gas flows.    The importance of the smaller cluster stars to the  structure of IRS2 may help understand the structure and evolution of other HII regions that hold embedded proto-clusters. 
  
 \section*{Acknowledgements}
Based on observations obtained at the international Gemini Observatory, a program of NSF’s NOIRLab, 
which is managed by the Association of Universities for Research in Astronomy (AURA) under a cooperative agreement 
with the National Science Foundation. on behalf of the Gemini Observatory partnership: the National Science Foundation
 (United States), National Research Council (Canada), Agencia Nacional de Investigaci\'{o}n y Desarrollo (Chile), 
 Ministerio de Ciencia, Tecnolog\'{i}a e Innovaci\'{o}n (Argentina), 
 Minist\'{e}rio da Ci\^{e}ncia, Tecnologia, Inova\c{c}\~{o}es e Comunica\c{c}\~{o}es (Brazil), and Korea Astronomy and Space Science Institute (Republic of Korea).

\section*{Data Availability}
The TEXES [NeII] data is available at the Gemini Observatory Archive under the Program ID GN-2006A-DS-2, and in FITS format at \url{https://github.com/danbeilis/data/tree/master/ngc7538}.  The radio continuum data may be found at the VLA Data Archive under Program AP374.



\bibliographystyle{mnras}
\bibliography{N7538} 

\begin{thebibliography}{}
\makeatletter
\relax
\def\mn@urlcharsother{\let\do\@makeother \do\$\do\&\do\#\do\^\do\_\do\%\do\~}
\def\mn@doi{\begingroup\mn@urlcharsother \@ifnextchar [ {\mn@doi@}
  {\mn@doi@[]}}
\def\mn@doi@[#1]#2{\def\@tempa{#1}\ifx\@tempa\@empty \href
  {http://dx.doi.org/#2} {doi:#2}\else \href {http://dx.doi.org/#2} {#1}\fi
  \endgroup}
\def\mn@eprint#1#2{\mn@eprint@#1:#2::\@nil}
\def\mn@eprint@arXiv#1{\href {http://arxiv.org/abs/#1} {{\tt arXiv:#1}}}
\def\mn@eprint@dblp#1{\href {http://dblp.uni-trier.de/rec/bibtex/#1.xml}
  {dblp:#1}}
\def\mn@eprint@#1:#2:#3:#4\@nil{\def\@tempa {#1}\def\@tempb {#2}\def\@tempc
  {#3}\ifx \@tempc \@empty \let \@tempc \@tempb \let \@tempb \@tempa \fi \ifx
  \@tempb \@empty \def\@tempb {arXiv}\fi \@ifundefined
  {mn@eprint@\@tempb}{\@tempb:\@tempc}{\expandafter \expandafter \csname
  mn@eprint@\@tempb\endcsname \expandafter{\@tempc}}}

\bibitem[\protect\citeauthoryear{Beilis, Beck  \& Lacy}{Beilis
  et~al.}{2021}]{Beilis_2021}
Beilis D.,  Beck S.,   Lacy J.,  2021, \mn@doi [MNRAS]
  {10.1093/mnras/stab3105}, 509, 2234

\bibitem[\protect\citeauthoryear{Bloomer et~al.,}{Bloomer
  et~al.}{1998}]{bloomer_98}
Bloomer J.~D.,  et~al., 1998, ApJ, 506, 727

\bibitem[\protect\citeauthoryear{Courant, Friedrichs  \& Lewy}{Courant
  et~al.}{1967}]{courant1967partial}
Courant R.,  Friedrichs K.,   Lewy H.,  1967, IBM J. Res. Dev., 11, 215

\bibitem[\protect\citeauthoryear{Ducati, Bevilacqua, Rembold  \&
  Riberio}{Ducati et~al.}{2001}]{ducati_2001}
Ducati J.~R.,  Bevilacqua C.,  Rembold S.,   Riberio D.,  2001, ApJ, 558, 309

\bibitem[\protect\citeauthoryear{Dullemond, Juhasz, Pohl, Sereshti, Shetty,
  Peters, Commercon  \& Flock}{Dullemond et~al.}{2012}]{dullemond2012radmc}
Dullemond C.,  Juhasz A.,  Pohl A.,  Sereshti F.,  Shetty R.,  Peters T.,
  Commercon B.,   Flock M.,  2012, ASCL, pp ascl--1202

\bibitem[\protect\citeauthoryear{Franco, Tenorio-Tagle  \& Bodenheimer}{Franco
  et~al.}{1990}]{franco1990formation}
Franco J.,  Tenorio-Tagle G.,   Bodenheimer P.,  1990, ApJ, 349, 126

\bibitem[\protect\citeauthoryear{Harten, Lax  \& Leer}{Harten
  et~al.}{1983}]{harten1983upstream}
Harten A.,  Lax P.~D.,   Leer B.~v.,  1983, SIAM Rev., 25, 35

\bibitem[\protect\citeauthoryear{{Henney}, {Arthur}  \&
  {Garc{\'\i}a-D{\'\i}az}}{{Henney} et~al.}{2005}]{henney_2005}
{Henney} W.~J.,  {Arthur} S.~J.,   {Garc{\'\i}a-D{\'\i}az} M.~T.,  2005,
  \mn@doi [\apj] {10.1086/430593}, \href
  {https://ui.adsabs.harvard.edu/abs/2005ApJ...627..813H} {627, 813}

\bibitem[\protect\citeauthoryear{Immer, Cyganowski, Reid  \& Menten}{Immer
  et~al.}{2014}]{Immer_2014}
Immer K.,  Cyganowski C.,  Reid M.~J.,   Menten K.~M.,  2014, A\&A, 563, 22

\bibitem[\protect\citeauthoryear{Kraus et~al.,}{Kraus
  et~al.}{2006}]{kraus_2006}
Kraus S.,  et~al., 2006, \mn@doi [A\&A] {10.1051/0004-6361:20065068}, 455, 521

\bibitem[\protect\citeauthoryear{Krti{\v{c}}ka}{Krti{\v{c}}ka}{2014}]{krtivcka2014mass}
Krti{\v{c}}ka J.,  2014, A\&A, 564, A70

\bibitem[\protect\citeauthoryear{Lacy, Richter, Greathouse, Jaffe  \& Zhu}{Lacy
  et~al.}{2002}]{lacy_2002}
Lacy J.,  Richter M.,  Greathouse T.,  Jaffe D.,   Zhu Q.,  2002, PASP, 114,
  153

\bibitem[\protect\citeauthoryear{Lamers \& Leitherer}{Lamers \&
  Leitherer}{1993}]{lamers1993mass}
Lamers H.~J.,  Leitherer C.,  1993, ApJ, 412, 771

\bibitem[\protect\citeauthoryear{Mignone}{Mignone}{2014}]{mignone2014high}
Mignone A.,  2014, J. Comput. Phys., 270, 784

\bibitem[\protect\citeauthoryear{Mignone, Bodo, Massaglia, Matsakos, Tesileanu,
  Zanni  \& Ferrari}{Mignone et~al.}{2007}]{mignone2007pluto}
Mignone A.,  Bodo G.,  Massaglia S.,  Matsakos T.,  Tesileanu O.,  Zanni C.,
  Ferrari A.,  2007, ApJS, 170, 228

\bibitem[\protect\citeauthoryear{Minier, Conway  \& Booth}{Minier
  et~al.}{2001}]{minier_2001}
Minier V.,  Conway J.~E.,   Booth R.~S.,  2001, A\&A, 369, 278

\bibitem[\protect\citeauthoryear{{Pirogov}}{{Pirogov}}{2009}]{pirogov_2009}
{Pirogov} L.~E.,  2009, \mn@doi [Astronomy Reports]
  {10.1134/S1063772909120051}, \href
  {https://ui.adsabs.harvard.edu/abs/2009ARep...53.1127P} {53, 1127}

\bibitem[\protect\citeauthoryear{Read}{Read}{1980}]{read_80}
Read P.,  1980, MNRAS, 193, 487

\bibitem[\protect\citeauthoryear{Sandell, Goss, Wright  \& Corder}{Sandell
  et~al.}{2009}]{Sandell_2009}
Sandell G.,  Goss W.~M.,  Wright M.,   Corder S.,  2009, \mn@doi [Astrophys.J.]
  {10.1088/0004-637X/699/1/L31}, 699, L31

\bibitem[\protect\citeauthoryear{Sandell, Wright, G{\"u}sten, Wiesemeyer,
  Reyes, Mookerjea  \& Corder}{Sandell et~al.}{2020}]{sandell_2020}
Sandell G.,  Wright M.,  G{\"u}sten R.,  Wiesemeyer H.,  Reyes N.,  Mookerjea
  B.,   Corder S.,  2020, ApJ, 904

\bibitem[\protect\citeauthoryear{Sato et~al.,}{Sato et~al.}{2010}]{sano_2010}
Sato J.,  et~al., 2010, ApJ, 724, 59

\bibitem[\protect\citeauthoryear{Scoville, Sargent, Sanders, Claussen, Masson,
  Lo  \& Phillips}{Scoville et~al.}{1986}]{scoville_1986}
Scoville N.,  Sargent A.,  Sanders D.,  Claussen M.,  Masson C.,  Lo K.~Y.,
  Phillips T.,  1986, Masers, Molecules and Mass Outflows in Star Forming
  Regions. Proceedings of a meeting held by the Haystack Observatory, Westford,
  Mass., USA, 15 - 16 May 1985.. A. D. Haschick (Editor). Haystack Observatory,
  Westford, Mass., USA. P. 201, 1986

\bibitem[\protect\citeauthoryear{Zhu, Lacy, Jaffe, Richter  \& Greathouse}{Zhu
  et~al.}{2008}]{zhu_2008}
Zhu Q.-F.,  Lacy J.~H.,  Jaffe D.~T.,  Richter M.~J.,   Greathouse T.~K.,
  2008, ApJS, 177, 584

\makeatother
\end{thebibliography}




\appendix

\onecolumn



\section{Line Profiles}

 \begin{figure}
\includegraphics[width=1\columnwidth]{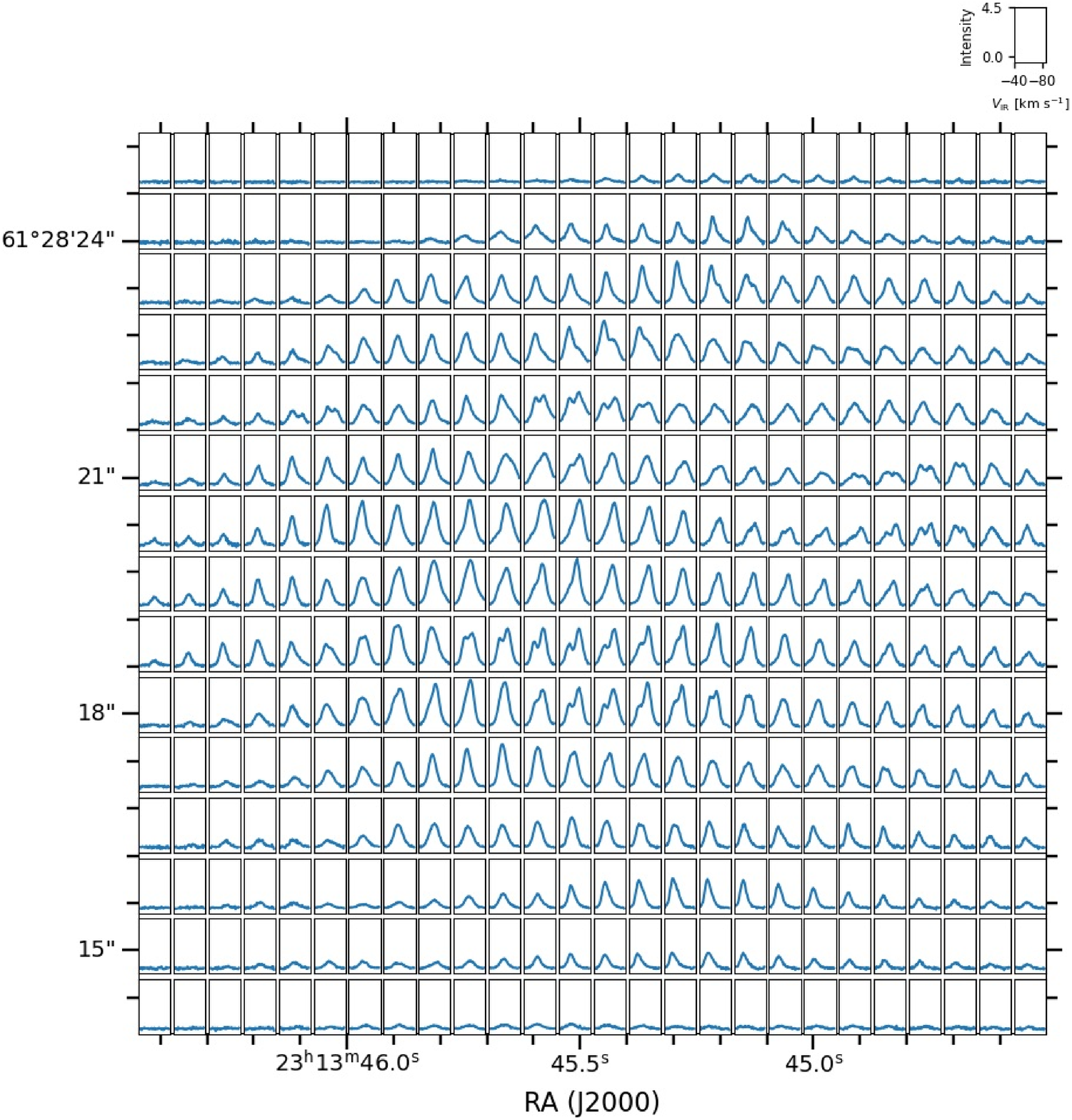}
		\caption{The line profile in 1 pixel at the given positions. The velocity scale and intensity scale are the same in all positions and are shown at the top right. The pixels shown are every 3rd pixel in Dec. and every 6th in R.A. }
\label{fig:profiles_figure}
\end{figure}

\twocolumn



\bsp	
\label{lastpage}
\end{document}